\documentclass{svproc}
\usepackage{url}

\usepackage{graphicx}

\newcommand{\bm}[1] {\mbox{\boldmath{$#1$}}}
\def\be{\begin{eqnarray}} 
 
\def\ee{\end{eqnarray}} 
\newcommand{\bi}{\begin{itemize}}
\newcommand{\ei}{\end{itemize}}
\def\sumint{\int \! \!\ \! \! \! \! \!\ \! \! \!\! \!\sum}
\def\Tr{\rm Tr}

\begin{document}
\mainmatter              
\title{Poincar\'e covariant light-front spectral function and transverse momentum distributions}
\titlerunning{Poincar\'e covariant light-front spectral function}  
%
\author{Emanuele Pace\inst{1}, Giovanni Salm\`e\inst{2} \and
Sergio Scopetta\inst{3}}
\authorrunning{Emanuele Pace et al.} 
%
%
\institute{ Universit\`a di Roma ``Tor Vergata'' and INFN, Sezione di  Roma Tor Vergata, 
Via della Ricerca Scientifica 1, 00133 Rome, Italy\\
\email{pace@roma2.infn.it}
\and
INFN, Sezione di Roma,  P.le A. Moro 2,
 I-00185 Rome, Italy\\
\and
 Universit\`a di Perugia and INFN, Sezione di Perugia, Via Alessandro Pascoli,
06123 Perugia, Italy}

\maketitle              
\vspace{-1mm}
\begin{abstract}
\vspace{-1mm}
In valence approximation the fermion correlator is simply related to the light-front spectral function. Then the leading twist  time-reversal even transverse momentum distributions can be explicitly  obtained from the  light-front wave function of the system and the twist-three distributions are linear combinations of the  transverse
distributions at leading twist. 
\keywords{Poincar\'e covariance, light-front Hamiltonian dynamics,  transverse momentum distributions}
\end{abstract}
\vspace{-4mm}
\section{Introduction}
\vspace{-1mm}
 Transverse momentum distributions (TMDs) are a powerful tool to study 
  hadron structure \cite{barone}. Light-cone models have been used to study the three-dimensional  hadron structure, to disentangle contributions from different  angular momentum components 
  and to investigate possible relations among the TMDs, with the aim to offer a guide for the extraction of TMDs from experimental data \cite{JMR,BP}. 
 
 In this paper  a
 Poincar\'e covariant, light-front (LF)
 spin-dependent spectral  function is considered to investigate hadrons within the LF Hamiltonian dynamics in valence approximation. We present both the most general expression for the spin-dependent momentum  distribution in terms of six scalar functions and
 a linear relation between the LF spectral function and the fermion correlator. This link implies approximate relations between the six time-reversal even (T-even) TMDs, as well relations between the leading twist and the twist-three TMDs \cite {noi2}.

The LF spectral  function, ${\cal P}^{\tau}_{{\cal M},\sigma'\sigma}({ \tilde {\bm \kappa}},\epsilon,S)$,
 was defined in Ref. \cite{noi}, starting from the LF wave function for a 
 three-body system with spin $1/2$, third component $\cal M$ and polarization vector $\bf S$. The energy $\epsilon$ is the energy  of a fully interacting {two-particle}  [23] subsystem and the variable $\tilde{\bm \kappa} = (\kappa^+, \bm \kappa_\perp)$ is the LF momentum for 
  particle 1 {in the intrinsic reference frame of the cluster [1,(23)]}. 
  The spectral function is defined through the overlaps between the LF wave function of the system  and the tensor product of a plane wave of 
  momentum $\tilde{\bm \kappa}$ and the state which describes the intrinsic motion of 
   the {two-particle spectator} subsystem.
The  mentioned tensor 
   product allows one to take care of macrocausality and  to introduce a new effect of binding in the spectral function. The LF spectral function, through the Bakamjian-Thomas construction of the Poincar\'e generators \cite{BT}, allows one to embed the successful phenomenology for few-nucleon systems in a Poincar\'e covariant framework and to satisfy at the same time normalization and momentum sum rule.
  {As a first test of our approach the EMC effect for $^3He$ is being evaluated. Preliminary results show encouraging improvements with respect to a  convolution approach with a momentum distribution \cite {EMC}.} 
   \vspace{-1mm}
       \section{ Light-Front spin-dependent spectral function and transverse momentum distributions}
 \subsection{Spin-dependent momentum distribution} 
 Integration of the LF spectral function
on the intrinsic energy {{$\epsilon$}} of the $(A-1)$ system, gives the LF spin-dependent momentum distribution  \cite{noi}
\be
n^\tau_{\sigma ' \sigma }(x,{\bf k}_{\perp};{\cal M}, {\bf S}) ~= ~ \sumint  d\epsilon ~{1 \over 2 ~ {(2\pi)^3}} ~ {1 \over 1 - x}  ~ {E_S \over \kappa^+} ~
{\cal P}^{\tau}_{{\cal M},\sigma'\sigma}({ \tilde {\bm \kappa}},\epsilon,S) \quad 
\quad \quad 
\ee 
where $\kappa^+ = x ~ {\cal M}_0[1,(23)]$, with ${\cal M}_0[1,(23)]$ the free mass of the cluster [1,(23)], and $E_S = \sqrt{4m^2 + 4 m \epsilon + |{\bm {\kappa}}|^2}$.
Within the LF approach, the 
momentum distribution 
  can be expressed through the three available independent vectors : i) the  polarization vector 
  {{$\bf S$}}; ii) the unit vector {{$\hat n$}} (identified with {{$\hat z$}}) which defines the {{$\pm$}} LF components, {{$v^\pm = v^0 \pm {\hat n} \cdot {\bf v}$}}, 
  and iii)
the transverse (with respect to the $z$ axis) momentum component
 {{${\bf k}_\perp = {\bf p}_\perp = \bm \kappa_\perp$}} of the momentum $\bf p$ 
 \be
 {{n^\tau_{\sigma ' \sigma }(x,{\bf k}_{\perp};{\cal M}, {\bf S}) 
 =
  {1 \over 2} \left\{ b_{0,{\cal M}} +
 {\bm \sigma} \cdot {\bm {f}_{{\cal M}}(x,{\bf k}_{\perp};\bf S)}\right\}_{\sigma ' \sigma } }}
 \label{nb1}
  \ee
  \vspace{-1mm}
   where ${{\bm f}}_{{\cal M}}(x,{\bf k}_{\perp};\bf S)$ is a pseudovector
   \vspace{-1mm}
  \be
 {\bm {f}}_{{\cal M}}(x,{\bf k}_{\perp};{\bf S}) &=& 
{\bf  S} ~
 b_{1,{\cal M}}
 ~
 + ~
\hat {\bf k}_{\perp} ~({\bf  S} \cdot \hat {\bf k}_{\perp})~
b_{2,{\cal M}}  
~ + ~ \hat {\bf k}_{\perp} ~({\bf  S} \cdot \hat z)~ 
b_{3,{\cal M}} 
\nonumber \\ 
&& + ~
 \hat z ~({\bf  S} \cdot \hat {\bf k}_{\perp})~
  b_{4,{\cal M}} ~
  + ~
 \hat z ~({\bf  S} \cdot \hat {z})~ 
 b_{5,{\cal M}}  \quad \quad.
\label{dv2}
\ee
 \hspace{-2mm} The functions {{$b_{i,{\cal M}}(x,{\bf k}_{\perp})$ }}$(i=0,1,...,5)$ and then 
 $n^\tau_{\sigma ' \sigma }(x,{\bf k}_{\perp};{\cal M}, {\bf S})$
  can be obtained from  the LF wave function of the system in momentum space. It results that
the spin-dependent momentum distribution is an 
integral on the relative intrinsic momentum $k_{23}$ of the interacting spectator pair \cite{noi2}
\vspace{-1mm}
 \be
\hspace{-1.cm} n^\tau_{\sigma  \sigma '}(x,{\bf k}_{\perp};{\cal M},{\bf S}) ~  
=  
~ {2 (-1)^{{\cal M} + 1/2} \over (1- x)}
 \int d { k}_{23}~
 \sum_L {\cal Z}^\tau_{\sigma \sigma'}(x,{\bf p}_{\perp}, k_{23}, L,{\bf S}) ~ 
~
\label{distr36}
\vspace{-3mm}
\ee
where $L$ is the orbital angular momentum 
of the contributions to 
the one-body off-diagonal density matrix
(only the values $L=0$ or $L=2$ are allowed).
\vspace{-1mm}
 \subsection{Fermion correlator and LF spectral function}
   The  fermion correlator in terms of the LF coordinates is \cite{barone}
     \be
{\Phi^{\tau}_{\alpha,\beta}(p,P,S)} ={ 1\over 2}\int {d\xi^- d\xi^+ d{\bm \xi}_T}~e^{ip \xi}
 \left \langle P,S,A|\bar{\psi}^{\tau}_{\beta}(0)\psi^{\tau}_{\alpha}(\xi)|A,S,P \right \rangle
\ee
where $|A,S,P  \rangle$ is the A-particle state
and $\psi^{\tau}_{\alpha}(\xi)$ the particle field (e.g. a nucleon of isospin ${\tau}$ in a nucleus, or 
a quark in a nucleon). 
A linear relation exists between the correlator in valence approximation, $\Phi^{\tau p}$, and the spectral function \cite{noi2}
   \be
\hspace{-.5cm}\Phi^{\tau p}_{\alpha,\beta}(p,P,S) ~
=~{ D \over 2 p^+ }
~\sum_{\sigma\sigma '}\left \{u_{\alpha}({\tilde {\bf p}},\sigma')~
{\cal P}^{\tau}_{{\cal M},\sigma'\sigma}({ \tilde {\bm \kappa}},\epsilon,S)~
{ \bar u}_\beta({\tilde {\bf p}},\sigma)\right \}  , ~
\label{abc}
\ee
where 
 $D= [(P^+)^2  ~ \pi ~ E_S ]/ \{ p^+ ~ m~ {\cal{M}}_0[1,(23)] \} $.
Then,
  traces of $\Phi^p$ 
can be expressed by traces of the spectral function :
\vspace{-1mm}
\be
\Tr(\gamma^+ \it{\Phi^p}) &=&
{{D}} ~
 Tr\left[ {\bm {\hat {\cal P}}}^{}_{\cal M}(\tilde {\bm \kappa},\epsilon,S)
\right] \quad \quad
\label{p}
\\
\Tr(\gamma^+ \gamma_5 ~ \it{\Phi^p}) 
&=& {{D}}~
Tr\left[ \sigma_z ~ {\bm {\hat {\cal P}}}^{}_{\cal M}(\tilde{\bm \kappa},\epsilon,S) \right] 
\quad
\label{p5}
\\ 
\Tr( \slash \! \! \! {\bf p}_\perp~\gamma^+ ~\gamma_5 ~ \it{\Phi^p}) 
&=& {{D}}
~
Tr\left[ {\bm p}_\perp \cdot {\bm \sigma} ~ {\bm {\hat {\cal P}}}^{}_{\cal M}(\tilde{\bm \kappa},\epsilon,S) \right] 
\label{ip5}
\vspace{-1mm}
\ee
The proper integration  on 
 $p^-$   
 of Eqs. (\ref{p},\ref{p5},\ref{ip5})
 and taking $p^+= xP^+$
 gives relations between the {{TMDs}} at leading twist and the functions 
 ~{{$b_{i,{\cal M}}$}} in Eqs. (\ref{nb1},\ref{dv2}), viz
\be
      {{ f(x, |{\bf p}_{\perp}|^2 ) }}
   = ~{{ {{b}}_{0}^{} }}    \quad  \quad  \quad  ~~
   &&    \quad  \quad  \quad
 {{ \Delta f(x, |{\bf p}_{\perp}|^2 ) }}
  = ~ {{{b}_{1,{\cal M}}^{}}} + ~ {{{b}_{5,{\cal M}}^{}}}
 \\  
  \hspace{-1.cm}  
 {{ g_{1T}(x, |{\bf p}_{\perp}|^2 )}}
= ~{{M} \over  |{\bf p}_{\perp}|} ~ {{{b}_{4,{\cal M}}^{}}}  
&& \quad  \quad
\hspace{-0.1cm}   ~~
{{\Delta'_T   f(x, |{\bf p}_{\perp}|^2 ) }}
= ~{ 1  \over 2} ~ 
\left \{  2 ~ {{{b}_{1,{\cal M}}^{} }}  +  
~
{{{b}_{2,{\cal M}}^{}  }} \right \}   \quad
\\  
{{h^{\perp}_{1L}(x, |{\bf p}_{\perp}|^2 )}}
= ~{M \over |{\bf p}_{\perp}|} ~ {{{b}_{3,{\cal M}}^{}}}  
&& \quad \quad \quad
{{h^{\perp}_{1T}(x, |{\bf p}_{\perp}|^2 )}}
   =  ~{ M^2 \over |{\bf p}_{\perp}|^2}   ~
{{{b}_{2,{\cal M}}^{}}}  ~
\vspace{-0.2cm}
\ee
\vspace{-1mm}
Linear equations between  transverse parton distributions were discussed in \cite{JMR} 
\vspace{-1mm}
\be
 \hspace{-1cm} \Delta f =  \Delta'_T   f ~ + ~ {|{\bf p}_{\perp}|^2 \over 2 M^2}~ h^{\perp}_{1T}\quad \quad \quad \quad 
 g_{1T} = - h^{\perp}_{1L}
\label{app3}
\ee
From the explicit expressions of the functions ${b}_{i,{\cal M}}$ in terms of the wave function of the system, one finds that these equalities hold exactly in valence approximation when the contribution to the transverse distributions from the angular momentum $L=2$ is absent. This implies a vanishing value of the orbital angular momentum of the particle in the system wave function \cite{noi2}.

On the contrary
the quadratic relation presented in \cite{JMR}  does not hold in our approach, even if  the contribution from the angular momentum $L=2$ is absent, because of the presence of $\int d k_{23}$ in the expressions (\ref{distr36}) of the transverse momentum distributions.

By evaluating proper traces of both the correlator and the spectral function, one can also obtain the {{twist-three}} TMDs in terms of the functions ${{b}_{i,{\cal M}}}$ and relate {{twist-three}} and  {{twist-two}} TMDs. In  our approximation  the same linear relations found in \cite{Bacchetta} hold, once the gluon contributions are disregarded. Obviously the T-odd TMDs vanish in  valence approximation \cite{noi2}.

\section{Conclusions and perspectives}
 The LF spin-dependent momentum distribution  for a spin $1/2$ system composed of three fermions (as  $^3He$ or a nucleon in valence approximation) can  be expressed through six functions 
 ~{{{${b}_{i,{\cal M}}$}}}, that can be written in terms of the LF wave function of the system in momentum space. 
A simple relation exists between the fermion correlator in valence approximation and the LF spectral function. Then it follows that the TMDs are combinations of the functions {{{${b}_{i,{\cal M}}$}}}.

As a result we found that the linear relations proposed between the T-even twist-two {{TMDs}}  hold in valence approximation 
{{whenever the contribution from the L=2 orbital angular momentum term in the one-body off-diagonal density matrix is absent}}, while the quadratic relation does not hold even in this case.
{Furthermore, in  valence approximation  the  proper relations between the {{twist-three}} and the  {{twist-two}} TMDs hold, once the gluon contributions are disregarded}.

{In the close future we will evaluate the transverse momentum distributions  for a nucleon in {{$^3He$}}, that could be extracted from {{measurements}} of appropriate {{spin asymmetries}} in   {{$^3{\overrightarrow{He}}({\overrightarrow {e}},e'p)$}} experiments at high momentum transfer. }



%
%

\end{document}